\documentclass[aps,prb,twocolumn,superscriptaddress]{revtex4}

\input epsf.sty

\begin{document}

\preprint{2/28/02 ver18}

%
%

\title{Ferroelectric Ordering in the Relaxor
Pb(Mg$_{1/3}$Nb$_{2/3}$)O$_{3}$ as Evidenced by Low-Temperature
Phonon Anomalies}

\author{S. Wakimoto} 
\email[Corresponding author: ]{waki@physics.utoronto.ca} \affiliation{
  Department of Physics, University of Toronto, Toronto, Ontario,
  Canada M5S~1A7 }

\author{C. Stock} 
\affiliation{ Department of Physics, University of Toronto, Toronto,
  Ontario, Canada M5S~1A7 }

\author{R. J. Birgeneau}
\affiliation{ Department of Physics, University of Toronto, Toronto,
  Ontario, Canada M5S~1A7 }

\author{Z.-G. Ye}
\affiliation{ Department of Chemistry, Simon Fraser University,
  Burnaby, British Columbia, Canada V5A~1S6 }

\author{W. Chen}
\affiliation{ Department of Chemistry, Simon Fraser University,
  Burnaby, British Columbia, Canada V5A~1S6 }

\author{W. J. L. Buyers}
\affiliation{ Neutron Program for Materials Research, National
  Research Council of Canada, Chalk River, Ontario, Canada K0J~1J0 }

\author{P. M. Gehring}
\affiliation{ NIST Center for Neutron Research, National Institute of
  Standards and Technology, Gaithersburg, MD 20899 }

\author{G. Shirane}
\affiliation{ Department of Physics, Brookhaven National Laboratory,
  Upton, NY 11973 }

\date{\today}

\begin{abstract}

%
%

  Neutron scattering measurements of the lowest-energy TO phonons in
  the relaxor Pb(Mg$_{1/3}$Nb$_{2/3}$)O$_3$ (PMN) are reported for $10
  \leq T \leq 750$~K.  The soft mode, which is overdamped by the polar
  nanoregions below the Burns temperature $T_d = 620$~K, surprisingly
  recovers below 220~K.  The square of the soft mode energy ($\hbar
  \omega_0$)$^2$ increases linearly with decreasing temperature, and
  is consistent with the behavior of a ferroelectric soft mode.  At
  10~K $\hbar \omega_0$ reaches 11~meV, the same value observed in
  ferroelectric Pb(Zn$_{1/3}$Nb$_{2/3}$)O$_3$ at low-$T$.  An unusual
  broadening of the TA phonon starts at $T_d$ and disappears at 220~K,
  coincident with the recovery of the TO mode.  These dynamics suggest
  that a well-developed ferroelectric state is established below
  220~K.

\end{abstract}


\maketitle

\section{Introduction}
%
%
%
%
Both Pb(Mg$_{1/3}$Nb$_{2/3}$)O$_3$ (PMN) and the Zn analogue (PZN) are
disordered perovskites of the form PbBO$_3$ that possess fascinating
dielectric properties.  Termed ``relaxors,'' each exhibits an unusual
dielectric response that peaks broadly at a temperature $T = T_{\rm
  max}$ that is strongly frequency dependent ($T_{\rm max} = 265$~K at
1 kHz for PMN).  When doped with ferroelectric PbTiO$_3$ (PT), both
compounds exhibit dramatic increases in their already exceptional
piezoelectric properties.~\cite{Park_JAP97} Each compound is
characterized by quenched chemical disorder on the perovskite B-site
by Mg$^{2+}$ (or Zn$^{2+}$) and Nb$^{5+}$ cations.~\cite{Ye_review} It
is well known that the mixed valence character inherent to both PMN
and PZN is required for the dielectric relaxation and ``diffuse
transition'' in a temperature range around $T_{\rm max}$.  Despite
these similarities, PMN and PZN are surprisingly different.  Whereas
PZN is a ferroelectric that transforms from a cubic to a rhombohedral
phase at 410~K,~\cite{Kuwata_82} PMN remains, on average, cubic below
$T_{\rm max}$ down to 5~K,~\cite{Mathan_91} To date the true ground
state of PMN remains an enigma.

%
%

%
%

%
%

Neutron measurements of the lattice dynamics and diffuse scattering in
these relaxor systems have proven to be invaluable in the effort to
solve the relaxor problem.  Pioneering work by Vakhrushev {\it et
  al.}~\cite{Vakhrushev_89,Vakhrushev_93,Naberezhnov_99} yielded
critical information about the diffuse scattering and low-frequency
transverse optic (TO) phonons in PMN at high temperatures.  Recent
neutron inelastic scattering measurements on PZN and PZN doped with PT
in the cubic phase have revealed an anomalous damping of the TO
phonons that only occurs at reduced wave vectors $q$ less than a
characteristic value $q_{wf} \sim 0.2$~\AA$^{-1}$, which gives rise to
the so-called ``waterfall'' feature.~\cite{Gehring_00_1,Gehring_00_2}
The origin of this damping was speculated to result from the presence
of local regions of randomly-oriented polarization that couple
strongly to the polar nature of TO modes.  These small regions of
polarization, also known as polar nanoregions or PNR, were first
inferred in 1983 by Burns and Dacol from measurements of the optic
index of refraction in different relaxor systems, including PMN and
PZN.~\cite{Burns_83} The most remarkable aspect of their finding was
that the PNR start forming at a temperature $T_d$, also known as the
Burns temperature, that is several hundred degrees higher than $T_{\rm
  max}$ (or $T_c$).

%
%

Gehring {\it et al.} demonstrated that at temperatures above $T_d$,
where there are no PNR, the low-$q$ TO modes in PMN exhibit a normal
optic phonon dispersion, and are not overdamped.~\cite{Gehring_01}
This provides the first direct evidence of the correlation between the
PNR and the anomalous phonon damping.  Subsequently, Hirota {\it et
  al.}~\cite{Hirota_01} proposed a simple model that resolves the
puzzling diffuse scattering problem in PMN by introducing the concept
of a ``phase-shifted condensed soft mode.''  Recent improvements in
crystal growth have made such measurements possible using larger
crystals.  From these experimental facts, we know that the PNR are
directly connected with the soft zone center TO mode.

In this paper we extend our previous phonon research down to 10~K
using a large high quality single crystal of PMN, the same as that
used in Ref.~[12].  Our results clearly show the recovery of the TO
mode at 220~K in addition to two distinct phonon anomalies which
provide definitive evidence of ferroelectric ordering in PMN, and thus
sheds new light on the ground state of PMN.

\begin{figure}
\centerline{\epsfxsize=3.4in\epsfbox{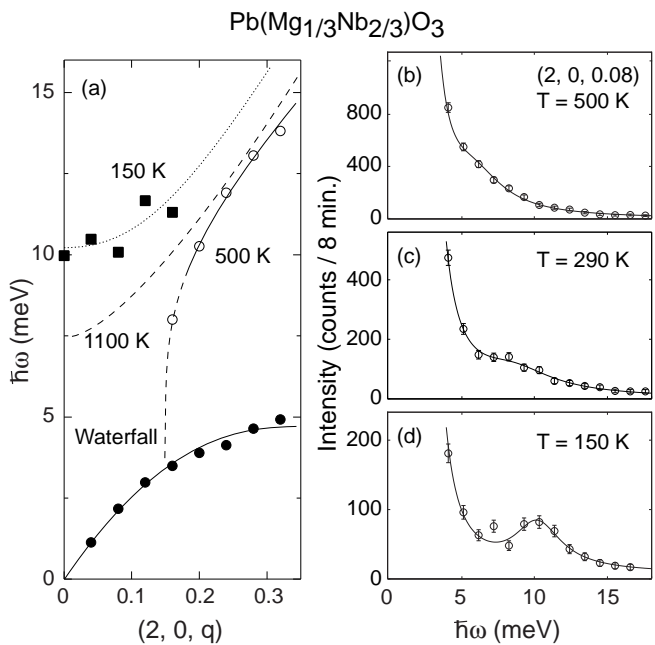}}
\caption{(a) TO phonon dispersion at 150~K (solid squares), 500~K 
  (open circles), and 1100~K (dashed line~\cite{Gehring_01}).  Solid
  circles show the TA phonon dispersion at 500~K.  Constant-$\vec{Q}$
  scan profiles at (2, 0, 0.08) are shown at (b) 500~K, (c) 290~K, and
  (d) 150~K.  Solid lines in (b), (c), and (d) are fits of the data to
  a resolution-convoluted Lorentzian function.  When compared to those
  at (2, 0, 0), these profiles suggest that higher-$q$ TO modes
  recover before those at lower-$q$ upon cooling below $T_c$. }
\end{figure}

\section{Experimental procedure}

Neutron-scattering experiments were performed on the C5
triple-axis spectrometer located in the NRU reactor at Chalk River
Laboratories.  Inelastic measurements were carried out by holding
the momentum transfer $\vec{Q} = \vec{k_i} - \vec{k_f}$ constant,
while scanning the energy transfer $\hbar \omega = E_i - E_f$ by
fixing the final neutron energy $E_f$ at 14.6~meV and varying the
incident energy $E_i$.  Horizontal beam collimations from reactor to
detector were 32$'$-32$'$-S-48$'$-60$'$ (S = sample).  Highly-oriented
pyrolytic graphite (HOPG) (002) crystals were used as monochromator,
analyzer, and higher-order filter.  A 3.25~g ($\sim 0.4$~cc) single
crystal of PMN was prepared by a top-seeded solution growth technique
from PbO flux.  The crystal was mounted in either a high-temperature
furnace or a closed-cycle helium refrigerator such that reflections of
the form $(h0l)$ lay in the scattering plane.  The room temperature
lattice parameter of PMN is $a=4.04$~\AA,~so that 1 reciprocal lattice
unit (r.l.u.)  $= 2\pi/a = 1.555$~\AA$^{-1}$.


\section{Results}
%
%
One of the most important findings of the present study is the
recovery of the soft mode at low temperature.  As shown in a previous
neutron scattering study, this mode becomes overdamped at $T_d$ when
the PNR first condense.~\cite{Gehring_01} The recovery of the TO
branch is summarized in Fig.~1(a) together with the TO branch at
1100~K from Ref.~[11].  Circles and squares represent the phonon
energies at 500~K and 150~K, respectively.  While the TO modes are
overdamped or heavily damped for $q \leq q_{wf} \sim 0.16$~r.l.u. at
500~K, these modes are underdamped at 150~K, and exhibit a normal TO
dispersion.  Phonon profiles at (2, 0, 0.08) ($q \le q_{wf}$) are
shown at three different temperatures in Fig.~1(b), (c), and (d).  It
is clear that the TO mode is nearly overdamped at 500~K, strongly
damped at 290~K, but has recovered an underdamped character at 150~K.

\begin{figure}
\centerline{\epsfxsize=2.8in\epsfbox{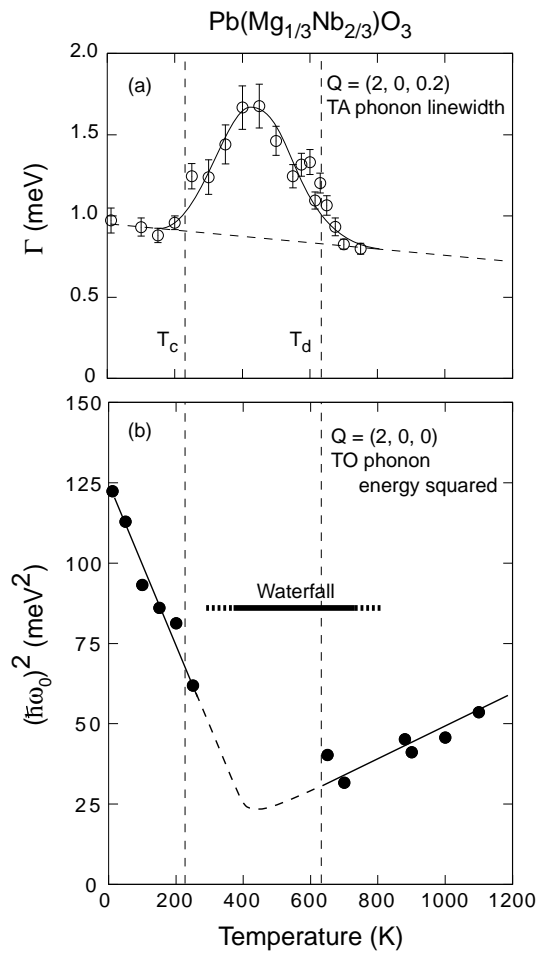}}
\caption{(a) TA phonon linewidth $\Gamma$ at (2, 0, 0.2) obtained
  from resolution-corrected fits to a Lorentzian function.  (b)
  Temperature dependence of the TO soft phonon energy squared.
  Vertical dashed lines correspond to $T_c=213$~K and $T_d=620$~K.
  The temperature range where the waterfall feature appears is
  indicated by the thick horizontal line. The other dashed and solid
  lines are guides to the eye.}
\end{figure}

%
Figure 2(a) summarizes the temperature dependence of the TA linewidth
$\Gamma$ at (2, 0, 0.2), which begins to broaden significantly near
$T_d$, and is in good agreement with the results of Naberezhnov {\it
  et al.}~\cite{Naberezhnov_99} and Koo {\it et al.}~\cite{Koo_CM}
Upon further cooling we obtain the new result that the broadening
peaks around 400~K and decreases until it finally disappears around
220~K.  Similar results are observed at $q=0.12$ and 0.16~r.l.u.,
however the broadening at $q=0.2$~r.l.u.\ was studied in more detail
since the effect is most prominent there.
%
Examples of TA phonon profiles at (2, 0, 0.2) are shown in the three
right-hand panels of Fig.~3.  Clearly the TA phonon linewidth at
500~K is larger than it is at either 150 or 750~K.  It is quite
interesting to note that the temperature at which the broadening
vanishes (220~K) is very close to the temperature at which PMN becomes
ferroelectric ($T_c = 213$~K) when cooled in the presence of a
sufficiently large electric field $E > E_c$.~\cite{Ye_review}

\begin{figure}
\centerline{\epsfxsize=3.4in\epsfbox{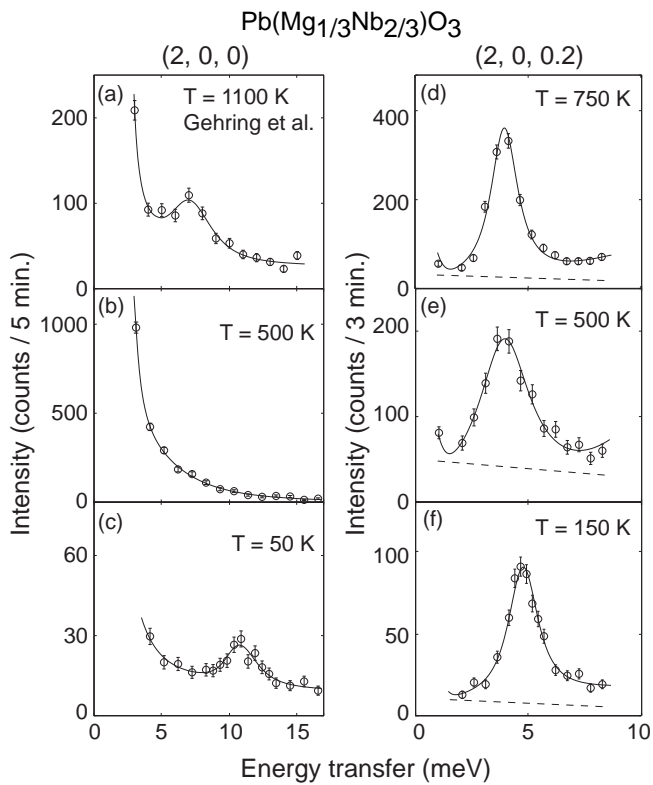}}
\caption{Temperature evolution of the TO phonon profile at (2, 0, 0)
  (left column) and TA phonon profile at (2, 0, 0.2).  Solid lines are
  fits to a resolution-convoluted Lorentzian function.  The data for
  1100~K at (2, 0, 0) by Gehring {\it et al.}~\cite{Gehring_01} were
  taken using a smaller (0.10~cm$^3$) crystal.}
\end{figure}

%
Fig.~2(b) combines the temperature dependence of the zone center TO
phonon energy squared ($\hbar\omega_0$)$^{2}$ measured at (2, 0, 0) at
low temperatures with that reported at high temperatures by Gehring
{\it et al.}~\cite{Gehring_01} When compared with Fig.~2(a) a truly
striking picture emerges.  As indicated by the dashed vertical lines
passing through both panels (a) and (b), the onset of the TA phonon
broadening near $T_d$ coincides with the temperature at which the zone
center mode becomes overdamped.  On the low temperature side, the TO
mode reappears around 220~K, which is effectively the same temperature
($\sim T_c$) at which the TA broadening vanishes.  Thus these data
show that the lattice dynamics of PMN are intimately connected with
the condensation of the PNR at $T_d$ as well as with the ferroelectric
transition temperature $T_c$ (for $E > E_c$).

At high temperatures ($T > T_d$), the zone center TO mode has been
shown to soften in a manner consistent with that of a ferroelectric
soft mode.~\cite{Gehring_01} It was not possible to determine the
energy $\hbar\omega_0$ of the zone center mode over a wide range of
temperatures below $T_d$ due to its overdamped nature (see Fig.~3(b)).
At temperatures below 220~K, however, the TO mode is no longer
overdamped, and the phonon profiles could be fit to a Lorentzian
function of $q$ and $\omega$ convolved with the instrumental
resolution function.  From these fits we have established that
($\hbar\omega_0$)$^2$ increases linearly with decreasing temperature.
This is a seminal point since, in a normal ferroelectric material, the
linear relationship between ($\hbar\omega_0$)$^2$ and the inverse
dielectric constant $1/\epsilon$ is well established above and below
$T_c$.~\cite{Shirane_70}
%
%
It should also be noted that $\hbar \omega_0$ reaches 11~meV at 10~K,
a value identical to that for ferroelectric PZN at low 
temperature.~\cite{PZN_TO}
This suggests a ferroelectric state is eventually established in PMN
too.  This point will be discussed in detail later.

\begin{figure}
\centerline{\epsfxsize=2.7in\epsfbox{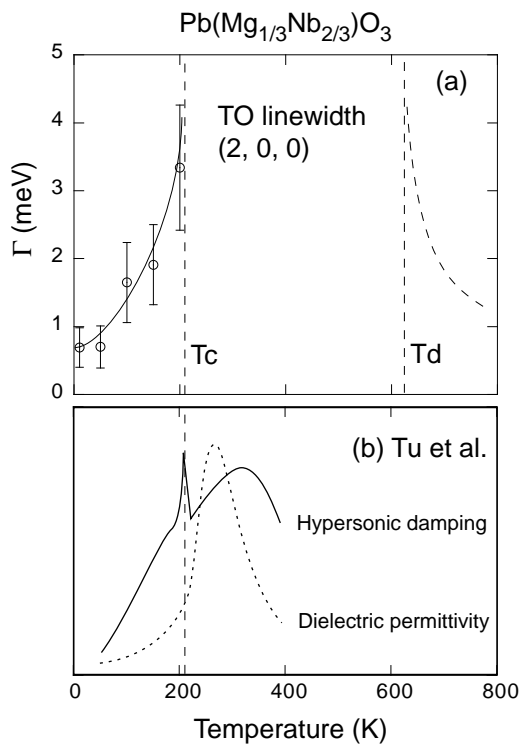}}
\caption{(a) Temperature dependence of the TO linewidth $\Gamma$ at
  (2, 0, 0).  (b) Schematic figure of the hypersonic damping (solid
  line) and dielectric constant (dotted line) versus temperature from
  Tu {\it et al.}~\cite{Tu_95} The vertical dashed line indicates
  $T=T_c$.}
\end{figure}

%
The three left-hand panels of Fig.~3 trace the evolution of the zone
center TO mode at (2, 0, 0).  We observe a well-defined propagating
mode at 1100~K, but a subsequently overdamped mode at 500~K, which
then completely recovers an underdamped character at 50~K.  The
dramatic recovery of the soft mode below 220~K occurs relatively
quickly with decreasing temperature.  By comparison, the recovery of
the underdamped soft mode in PZN is slower, as it is readily observed
only at temperatures below $\approx 100$~K.~\cite{PZN_TO}
%
%
The quick recovery in PMN is also clearly demonstrated in Fig. 4(a),
which shows the temperature dependence of the soft mode linewidth
$\Gamma$.  Below $T_c$ $\Gamma$ decreases quickly, and becomes very
small at low temperature.

\section{Discussion}

In Fig.~2 we demonstrate the occurence of two distinct phonon
anomalies in PMN near 220~K.  The most remarkable of these is the
linear increase of the square of the soft mode energy that begins
below 220~K.  This behavior is typical of many ferroelectrics in their
ordered states.  Yet PMN is known to retain an average cubic structure
through 220~K, and even down to very low temperature.~\cite{Mathan_91}
The soft mode behavior shown in Fig.~2 appears to contradict the
currently accepted picture of PMN because it implies that a phase
transition takes place from a relaxor state to a short-range ordered
ferroelectric (non-cubic) state at $T_c$.

Relaxor-to-ferroelectric phase transitions have been observed in other
lead-oxide compounds such as chemically disordered
Pb(Sc$_{1/2}$Ta$_{1/2}$)O$_3$ (PST)~\cite{Chu_93} and
Pb(Sc$_{1/2}$Nb$_{1/2}$)O$_3$ (PSN).~\cite{Chu_95} They usually occur
at a temperature $T_c$ that is lower than the temperature of the
maximum permittivity $T_{\rm max}$.  The difference between $T_{\rm
  max}$ (at 1~kHz) and $T_c$ is about 15 degrees in PST and PSN, and
is significantly smaller than that observed in PMN ($\approx 50$~K).
In addition, both PST and PSN have relatively sharp dielectric
permittivity $\epsilon(T)$ peaks (small diffuse character) compared to
that in PMN.  The only data on PMN that shows an anomaly at $T_c =
213$~K upon ZFC is the peak in the hypersonic damping shown in
Fig.~4(b).~\cite{Tu_95} It has also been reported that a macroscopic
ferroelectric phase can be induced in PMN upon cooling in a small
electric field, and that once induced, the ferroelectric phase and the
related polarization vanish upon zero-field-heating as a first-order
transition at $T_c=213$~K.~\cite{Ye_93,Calvarin_95} These results
suggest that PMN may exhibit a ferroelectric phase, which, however,
does not achieve long-range order when zero-field cooled.  Therefore,
the soft mode behavior reported here is not entirely unexpected.  It
does, however, reveal the ferroelectric phase transition near $T_c$ in
terms of the dynamics for the first time.

There remains the question of why a sharp structural phase transition
is not observed in PMN by x-rays, even though a ferroelectric phase is
implied from the low-temperature lattice dynamics.  One possible
answer concerns the length scale of the rhombohedral ferroelectric
phase.  Quite simply, a length scale of order $\sim100$~\AA\ will
appear as long-range order to neutrons but as short-range order to
x-rays.  In fact, ferroelectric regions of order 100~\AA\ in size have
been reported by de Mathan {\it et al.}~\cite{Mathan_91}

PMN is a special relaxor compound given its unusually broad dielectric
relaxation, and the large temperature difference between $T_{\rm max}$
and $T_c$.  We believe that an important clue to understanding the low
temperature properties of PMN lies in the concept of the phase-shifted
condensed soft mode recently proposed by Hirota {\it et
  al.}~\cite{Hirota_01} In this work the PNR are shown to result from
the condensation of the soft TO mode at $T_d$ through a model that
reconciles the differences between the structure factors of the soft
mode and the diffuse scattering by introducing a uniform ionic shift
of the PNR along their polar axis relative to the surrounding cubic
matrix.  This phase shift persists to low temperatures as shown by the
structural work of de Mathan {\it et al.}~\cite{Mathan_91} It is
important to note that the relative intensities of the diffuse lines
reported by de Mathan {\it et al.} are in excellent agreement with the
phase-shifted condensed soft mode results of Hirota {\it et al.} The
uniform phase shift idea provides an additional energy barrier to the
formation of a long-range ordered ferroelectric state.  We speculate
that the combination of this energy barrier and the random field
effects introduced by the PNR prevent PMN from entering into a pure
ferroelectric state.

Finally, we discuss the relationship between $(\hbar\omega_{0})^2$ and
$1/\epsilon(T)$.  In Fig. 2(b), the dashed line is drawn so that
$(\hbar\omega_{0})^2$ has a minimum at $\sim 400$~K, where the TA
phonon linewidth is largest.  Since $(\hbar\omega_{0})^2$ is linearly
related to $1/\epsilon(T)$, one might guess that the minimum value of
$(\hbar\omega_{0})^2$ would occur at $T_{max} = 265$~K.  However, this
value of $T_{max}$ is obtained at a measuring frequency of 1~kHz.  By
comparison, $T_{\rm max} = 305$~K at a measuring frequency of
100~MHz.~\cite{Tsurumi}.  Our neutron measurements probe the dynamical
response function of PMN at frequencies on the order of THz.  Since
$T_{max}$ is by definition strongly frequency dependent in relaxors,
the temperature of 400~K may possibly correspond to $T_{max}$ on the
THz scale.

To understand the relaxor-to-ferroelectric phase transition in PMN
(and thos in other relaxors), we intend to carry out a series of
neutron experiments which includes a systematic comparison of the
lattice dynamics in PMN, PMN-$x$PT, PSN and PZN. These experiments
will provide useful information about the microscopic mechanism of
relaxor ferroelectricity.  Recent studies~\cite{Koo_CM,Dkhill_02} of
PMN-$x$PT have already documented very interesting changes in the
phase transition with increasing PbTiO$_3$ content $x$.

\begin{acknowledgments}
  We thank A.\ A.\ Bokov, K.\ Hirota, J.\ -M.\ Kiat, B.\ Noheda, and
  K.\ Ohwada for stimulating discussions.  Work at the University of
  Toronto is supported by the Natural Science and Engineering Research
  Council of Canada.  We acknowledge financial support from the U.\ 
  S.\ DOE under contract No.\ DE-AC02-98CH10886, and the Office of
  Naval Research under Grant No.\ N00014-99-1-0738.
\end{acknowledgments}


\begin{references}

\bibitem{Park_JAP97} S.-E. Park and T. R. Shrout, J. Appl. Phys. {\bf
    82}, 1804 (1997).

\bibitem{Ye_review} See review article, Z.-G. Ye, {\it Key Engineering
    Materials Vols. 155-156}, 81 (1998).

\bibitem{Kuwata_82} J. Kuwata, K. Uchino, and S. Nomura,
  Ferroelectrics {\bf 37}, 579 (1981); Jpn J. Appl. Phys. {\bf 21},
  1298 (1982).

\bibitem{Mathan_91} N de Mathan, E. Husson, G. Calvarin, J. R.
  Gavarri, A.  W. Huwat, and A. Morell, J. Phys. Condens. Matter {\bf
    3}, 8159 (1991).

\bibitem{Vakhrushev_89} S. B. Vakhrushev, B. E. Kvyatkovksy, A. A.
  Naberezhnov, N. M. Okuneva, and B. Toperverg, Ferroelectrics {\bf
    90}, 173 (1989).

\bibitem{Vakhrushev_93} S. B. Vakhrushev, A. A. Naberezhnov, N. M.
  Okuneva, and B. N. Savenko, Phys. Solid State {\bf 37} 1993 (1995).

\bibitem{Naberezhnov_99} A. Naberezhnov, S. Vakhrushev, B. Doner, D.
  Strauch, and H.  Moudden Eur. Phys. J. B {\bf 11}, 13 (1999)

\bibitem{Gehring_00_1} P. M. Gehring, S.-E. Park, and G. Shirane,
  Phys. Rev. Lett. {\bf 84}, 5216 (2000).

\bibitem{Gehring_00_2} P. M. Gehring, S.-E. Park, and G. Shirane,
  Phys. Rev. B {\bf 63}, 224109 (2000).

\bibitem{Burns_83} G. Burns and F. H. Dacol, Solid State Commun. {\bf
    48}, 853 (1983).

\bibitem{Gehring_01} P. M. Gehring, S. Wakimoto, Z.-G. Ye, and G.
  Shirane, Phys. Rev. Lett. {\bf 87}, 277601 (2001).

\bibitem{Hirota_01} K. Hirota, Z.-G. Ye, S. Wakimoto, P. M. Gehring,
  and G.  Shirane, Phys. Rev. B {\bf 65}, 104105 (2002).

\bibitem{Koo_CM} T.-Y. Koo, P. M. Gehring, G. Shirane, V. Kiryukhin,
  G. Lee, and S.-W. Cheong, Phys. Rev. B (in press), cond-mat/0110531.

\bibitem{Shirane_70} G. Shirane, J. D. Axe, and J. Harada, Phys. Rev.
  B {\bf 2}, 155 (1970).

\bibitem{PZN_TO} P. M. Gehring, unpublished.

\bibitem{Chu_93} F. Chu, N. Stter and A. K. Tagantsev, J. Appl. Phys.,
  74, 5129 (1993).

\bibitem{Chu_95} F. Chu, I. M. Reaney, and N. Stter, J. Appl. Phys.,
  77, 1671 (1995).

\bibitem{Tu_95} C.-S. Tu, V. Hugo Schmidt, and I. G. Siny, J. Appl.
  Phys. {\bf 78}, 5665 (1995).

\bibitem{Ye_93} Z.-G. Ye and H. Schmid, Ferroelectrics {\bf 145}, 83
  (1993).

\bibitem{Calvarin_95} G. Calvarin, E. Husson, and Z.-G. Ye,
  Ferroelectrics {\bf 165}, 349 (1995).

\bibitem{Tsurumi} T. Tsurumi, K. Soejima, T. Kamiya, and M. Daimon,
  Jpn. J. Appl. {\bf 33}, 1959 (1994).

\bibitem{Dkhill_02} B. Dkhil, J. M. Kiat, G. Calvarin, G. Baldinozzi,
  S. B.  Vakhrushev, and E. Suard, Phys. Rev. B {\bf 65}, 024104
  (2002).

\end{references}
\end{document}